\documentclass[journal]{IEEEtran}
\usepackage{cite}
\usepackage{amsmath,amssymb,amsfonts}
\usepackage{algorithmic}
\usepackage{graphicx}
\usepackage{textcomp}
\def\BibTeX{{\rm B\kern-.05em{\sc i\kern-.025em b}\kern-.08em
    T\kern-.1667em\lower.7ex\hbox{E}\kern-.125emX}}
\begin{document}

\title{Transient Stability Analysis of Grid-Connected Converters Based on Reverse-Time Trajectory}
\author{Mohammad Kazem Bakhshizadeh, Sujay~Ghosh, Guangya Yang and~Łukasz Kocewiak
\thanks{M. K. Bakhshizadeh, S. Ghosh and Ł. Kocewiak are with Ørsted Wind Power,
Nesa Allé 1, 2820, Denmark. E-mail: sujgh@orsted.com}
\thanks{G. Yang is with the Technical University of Denmark, Anker Engelunds Vej 1, 2800, Denmark}
\thanks{Manuscript received March xx, 2023}}
\maketitle

\begin{abstract}
As the proportion of converter-interfaced renewable energy resources in the power system is increasing, the strength of the power grid at the connection point of wind turbine generators (WTGs) is gradually weakening. Existing research has shown that when connected with the weak grid, the dynamic characteristics of the traditional grid-following controlled converters will deteriorate, and unstable phenomena such as oscillation are prone to arise. Due to the limitations of linear analysis that can not sufficiently capture the stability phenomena, transient stability must also be investigated. So far, standalone time-domain simulations or analytical Lyapunov stability criteria have been used to investigate transient stability. However, time-domain simulations have proven to be computationally too heavy, while analytical methods are more complex to formulate, require many assumptions, and are conservative. This paper demonstrates an innovative approach to estimating the system boundaries via hybrid - linearised Lyapunov function-based approach and the time-reversal technique. The proposed methodology enables compensation for both time-consuming simulations and the conservative nature of Lyapunov functions. This work brings out the clear distinction between the system boundaries with different post-fault active current ramp rate controls. At the same time providing a new perspective on critical clearing times for wind turbine systems. Finally, the stability boundary is verified using time domain simulation studies.
\end{abstract}

\begin{IEEEkeywords}
Lyapunov direct method, Non-autonomous systems, PLL, Time trajectory reversal, Transient stability assessment, Wind turbine converter system.
\end{IEEEkeywords}

\section*{Nomenclature}
\normalsize
\renewcommand{\arraystretch}{1.3}
\begin{tabular}{p{1.7cm}p{6.0cm}}
$\mathbb{R}^n$ & Is a n dimensional Euclidean space. \\
0$^n$ & Is a n×1 null vector. \\
$\in$ & Is a member of. \\
$\subset$  & Is a subset of. \\
$\partial \mathbb{A}$ & Boundary of a space A. \\ 
$\forall$ & For all. \\
$\exists$ & There exists. \\
$A \Rightarrow B$ & A results in B. A is a sufficient condition for B. B is a necessary condition for A. \\
$A \Leftrightarrow B$ & B is true/false if and only if A is true/false. A/B is a sufficient and necessary condition for B/A. 
\end{tabular}

\normalsize
\renewcommand{\arraystretch}{1.3}
\begin{tabular}{p{2.4cm}p{5.3cm}}
$\textbf{\textit{x}}^T$ & Transpose of vector $\textbf{\textit{x}}$. \\
$f: X\rightarrow Y$ & Function $f$ that maps set $X$ to set $Y$. \\
$f^{-1}(\textbf{\textit{x}})$ & Inverse of function $f(\textbf{\textit{x}})$. \\
$\dot{\textbf{\textit{x}}}=\frac{d\textbf{\textit{x}}}{dt}$ & Time derivative of vector $\textbf{\textit{x}}(t)$. \\
$\textbf{A}|_{\textbf{\textit{x}}=\textbf{\textit{x}}_0}$ & $\textbf{A}$ is evaluated at $\textbf{\textit{x}}=\textbf{\textit{x}}_0$. \\
$\Re\{Z\}$ & Real part of the complex number $Z$. \\
$\lambda(\textbf{A})$ & Set of the eigenvalues of matrix \textbf{A}. \\
$\textbf{\textit{x}}_{ini}$  & Initial condition for a dynamical system. \\
$\textbf{\textit{x}}_{final}$  & Final condition for a dynamical system. \\
$\textbf{\textit{x}}(t)= \varphi(t, \textbf{\textit{x}}_{ini})$ & Solution of a dynamical system for a specific initial condition. \\ 
$
\frac{\partial f}{\partial x} =
\begin{bmatrix}
  \frac{\partial f_1}{\partial x_1} &     \cdots & 
    \frac{\partial f_1}{\partial x_n} \\ 
  \vdots &     \ddots &     \vdots \\
  \frac{\partial f_n}{\partial x_1} &     \cdots & 
    \frac{\partial f_n}{\partial x_n}
\end{bmatrix} $ & Jacobian of the vectorial function $f$. \\
$\textbf{P}	\succ 0$ & Positive-definite matrix \textbf{P}. \\
$\textbf{P}	\succeq 0$ & Positive-semidefinite matrix \textbf{P}. \\
$\Tilde{\textbf{\textit{x}}}$ & Equilibrium point/state of a dynamical system. \\
$F_{abc} =
\begin{bmatrix}
  F_a \\ 
  F_b \\
  F_c      
\end{bmatrix}$ & Representation of signal F(t) in the three-phase stationary frame (abc domain). \\
$F_{dq} =
\begin{bmatrix}
  F_d \\ 
  F_q 
\end{bmatrix}$ & Representation of signal F(t) in the rotating frame (dq domain). 
\end{tabular}

\section{Introduction}
\label{sec:introduction}
\IEEEPARstart{A}{s} of 2021, the worldwide installation of wind power capacity has reached approximately 743 GW, contributing significantly to a reduction of over 1.1 billion tonnes of CO2 emissions globally \cite{1}. The wind industry is poised for continued growth due to technological innovations, economies of scale, and policy support around the world. However, the increasing proportion of converter-interfaced renewable energy resources in the power system \cite{2} has weakened the connection strength between wind turbine generators (WTGs) and the power grid. Previous research has demonstrated that the dynamic characteristics of traditional grid-following controlled converters can deteriorate when connected to a weak grid, leading to unstable phenomena such as oscillation \cite{3}\cite{4}.

Traditionally, the stability of wind farm connections has been analysed using linearised model-based approaches, such as eigenvalue analysis \cite{7}\cite{8} or impedance-based stability analysis \cite{9}\cite{11}. These methods assume that the system, including the wind turbine (WT) and the connected power system, behaves linearly under small disturbances and that stability is only analysed within the operating point's vicinity. However, it has been noted in \cite{12} that small-signal stability assessment alone cannot guarantee overall stability. Therefore, transient stability must also be investigated to ensure that the system remains stable under larger disturbances.

In \cite{13}, it has been demonstrated that large disturbances can destabilise the phase-locked loop (PLL), which can have a significant impact on the transient stability of the wind turbine (WT) system. In general, transient stability has been evaluated through time-domain simulations. Time-domain simulation is simple, however, it cannot provide a close-form solution for quantifying stability margins. Therefore, it is necessary to repeat the simulations over a large set of system conditions (like phase portraits) to identify the system boundary, i.e. the region of attraction (RoA) \cite{14}.

Alternatively, analytical transient stability methods, such as equal area criteria and Lyapunov's direct method \cite{15}, provides a closed-form solution for the system. Here an non-linear energy function is constructed such that after a disturbance the decrease in energy results in a stable system.
A classical non-linear energy function is constructed for synchronous generators based on its swing equation \cite{16}. Efforts have been made to extend the same analysis to WT systems \cite{17}; however, the system is assumed to have autonomous behaviour, see Section III. In \cite{18}, a non-linear energy function for a WT with non-autonomous behaviour is constructed based on \cite{19}, which states that a system has a smaller RoA when the active current ramp post-fault is faster. However, the approach to construct the energy function is highly complex and results in a conservative estimate of the RoA.

Recent developments have focused on maximising the system's RoA by formulating an optimisation problem using sum-of-squares programming \cite{20}. Additionally, some machine learning (ML) techniques \cite{22} have been studied to achieve a better estimate of the RoA. However, these methods require significant expertise in data-driven techniques. 
Considering, (a) the high computation burden of repeated time domain simulations over a large set of system conditions, (b) the mathematical complexities of non-linear analytical methods coupled with conservative estimate of system RoA, and (c) the requirement of domain expertise in data-driven techniques for optimisation and ML methods; the objective of this paper is to propose a fast and simplified transient stability assessment method that can be easily adopted by the industry. 

This paper presents a novel approach to transient stability assessment of wind power plants (WPPs) by combining the advantages of time-domain simulations and analytical energy-based stability methods. Specifically, we use the reverse time trajectory technique in conjunction with linear Lyapunov functions to estimate the system boundary. Compared to nonlinear energy functions, the construction of linear energy function is simple and has an established procedure. Additionally, the reverse time simulation only needs to be performed for stable cases, significantly reducing the number of repeated time domain simulations.

The time-reversal technique has been the subject of extensive research for several decades \cite{24}-\cite{26}. The application of time-reversal in dynamic systems dates back to 1915, where it was initially used to analyse a three-body problem \cite{27}. Subsequently, time-reversal has been employed in various problems related to thermodynamics and quantum mechanics, as discussed in \cite{28}-\cite{30}. Reference \cite{31} provides a extensive overview of time reversible dynamics, including system equations, and conservative and dissipative behavior.
Building on our previous research on nonlinear modelling and transient stability assessment of WPPs \cite{18},\cite{33}-\cite{35}. Our proposed methodology aims to provide a fast, simple and practical solution for industry without requiring complex mathematical analysis. The following is the contribution in the paper,

\begin{enumerate}
    \item A hybrid approach to estimating the post-fault system boundary (RoA) is proposed based on linear energy function and reverse time-trajectory. 
    
    \item This work brings out the clear distinction between the system boundaries with different post-fault active current ramp rate controls.
    
    \item A new perspective on critical clearing time for wind turbine systems is discussed. 
\end{enumerate}

Section II provides an overview of the mathematical preliminaries for the proposed transient stability assessment method. Section III details the large signal reduced order WT model and its transient stability assessment. The time-domain validation of the proposed method is presented in Section IV. The paper concludes with Section V.

\section{Transient Stability - Mathematical Preliminaries}
Most of the dynamical systems can be described by the following ordinary differential equation (ODE):

\begin{equation}\label{eq:1}
      \dot{x} = f(t, x, u)\\
\end{equation}
where $t$ is time, $\dot{\textbf{\textit{x}}}$ is the time derivative of vector $\textbf{\textit{x}}\in\mathbb{R}^n$, and $\textbf{\textit{u}}\in\mathbb{R}^m$ is the vector of input signals. $\textbf{\textit{x}}$ is called the state variables of the system. Usually, the inputs are defined based on time and the state variables, therefore, they can be omitted from (\ref{eq:1}). If $f$ is not an explicit function of time, then the system defined by (\ref{eq:2}) is called an autonomous system \cite{Khalil}.

\begin{equation}\label{eq:2}
      \dot{x} = f(x)\\
\end{equation}

An equilibrium point for a dynamical system is defined as a point $\Tilde{\textbf{\textit{x}}}$, for which $f(\Tilde{\textbf{\textit{x}}})=0$. In other words, if the system solution $\textbf{\textit{x}}(t)$ reaches $\Tilde{\textbf{\textit{x}}}$ it stays there forever.

\subsection{Lyapunov's direct method for stability analysis}
Scalar continuous and differentiable function $V:D \subset \mathbb{R}^n \rightarrow \mathbb{R}$ is called a Lyapunov function (LF) for the system (\ref{eq:2}) with $\Tilde{\textbf{\textit{x}}}=0$ such that,

\begin{itemize}
    \item $V(0^n) = 0$ $\Leftrightarrow$ $\textbf{\textit{x}} =  0^n$ \\
    \item $V(\textbf{\textit{x}}) > 0$ $\forall$ $\textbf{\textit{x}} \in \mathbb{D}-\{0^n\}$ \\
    \item $\dot{V}(\textbf{\textit{x}}) \leq 0$  $\forall$ $\textbf{\textit{x}} \in \mathbb{D}$
\end{itemize}
which also shows that the system is stable. Furthermore, if $\dot{V}(\textbf{\textit{x}})<0$ $\forall \textbf{\textit{x}} \in \mathbb{D} - \{0^n\}$, then the system is asymptotically stable, i.e. $\lim_{t \to \infty} \textbf{\textit{x}}(t) = 0^n$. It must be noted that if the equilibrium point $\Tilde{\textbf{\textit{x}}}$ is not the origin, it can be shifted by change of variables.

\subsection{Region of Attraction for dynamical systems}
The region of attraction for an equilibrium point is defined as a set,

\begin{equation}\label{eq:3}
      \mathbb{D} = \{ \textbf{\textit{x}}_{init} \in \mathbb{R}^n : \lim_{t \to \infty} \varphi(t, \textbf{\textit{x}}_{init}) = 0^n\}\\
\end{equation}

Finding the exact RoA is a highly complex task; instead, finding an inner estimate of the exact RoA is common practice. A set defined by $V(\textbf{\textit{x}}) \leq c$  ($c>0$) is called a sublevel set of the LF $V(\textbf{\textit{x}})$, which is a set that if the solution trajectory $\textbf{\textit{x}}(t)$ enters, then it cannot exit. Therefore,

\begin{equation}\label{eq:4}
      V(x)\leq c \subset \mathbb{D}\\
\end{equation}

In other words, obtaining the biggest estimate of the RoA is to find appropriate LFs and then maximize c.

\begin{figure}[h]
    \centering
    \includegraphics[width=7.5cm]{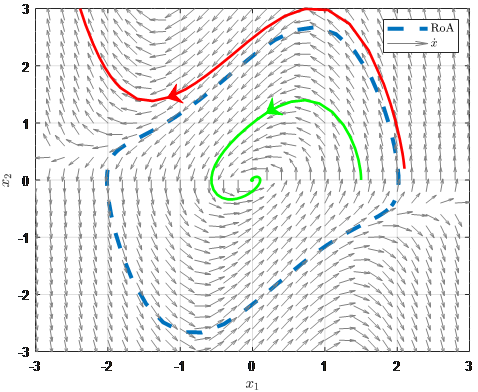}
    \caption{RoA of the reversed Van der Pol system (\ref{eq:5}).}
    \label{fig:Rvanderpol}
\end{figure}

For example, the RoA of the reversed Van der Pol system (\ref{eq:5}) is presented in Fig. \ref{fig:Rvanderpol}, where it is evident that if the initial point is inside the RoA, the system is attracted to the origin.

\begin{equation}\label{eq:5}
    \begin{cases}
      \dot{x}_1 = -x_2\\
      \dot{x}_2 = x_1 - x_2(1-x_1^2)
    \end{cases}       
\end{equation}

\subsection{Lyapunov function candidate from linerised system}
The nonlinear dynamical system (2) can be approximated by a linear model in a small region around the operating point (i.e. origin) by small signal linearisation as follow,

\begin{equation}\label{eq:7}
      \Delta \dot{\textbf{\textit{x}}} = \textbf{A} \Delta \textbf{\textit{x}}\\
\end{equation}
where $\textbf{A} = \frac{\partial f}{\partial x}|_{\textbf{\textit{x}}=0^n}$

If \textbf{A} is Hurwitz matrix, then a quadratic LF $V(\textbf{\textit{x}})$ can easily be found by using the linearised model (\ref{eq:7}) as,

\begin{equation}\label{eq:8}
      V(\textbf{\textit{x}}) = \textbf{\textit{x}}^T\textbf{P}\textbf{\textit{x}}\\
\end{equation}
where, for any $\textbf{Q}\succ0$, $\textbf{P}\succ0$ is the solution of the Lyapunov equation,
\begin{equation}\label{eq:9}
      \textbf{P}\textbf{A} + \textbf{A}^T\textbf{P} + \textbf{Q} = 0\\
\end{equation}

For example, for the reversed Van der Pol system (\ref{eq:5}), the linearisation around the origin results in, 

\begin{equation}\label{eq:10}
      \textbf{A} =
\begin{bmatrix}
  0 &     -1 \\ 
  1 &    -1
\end{bmatrix} 
\end{equation}
which is Hurwitz. By assuming $\textbf{Q} = \begin{bmatrix}
  1 &     -0.5 \\ 
  -0.5 &    1
\end{bmatrix} $ $\succ$ 0 and solving the Lyapunov equation in (\ref{eq:9}) results in,

\begin{equation}\label{eq:11}
      \textbf{P} =
\begin{bmatrix}
  1 &    -0.5 \\ 
  -0.5 &    1
\end{bmatrix} 
\end{equation}

\begin{figure}[h]
    \centering
    \includegraphics[width=7.5cm]{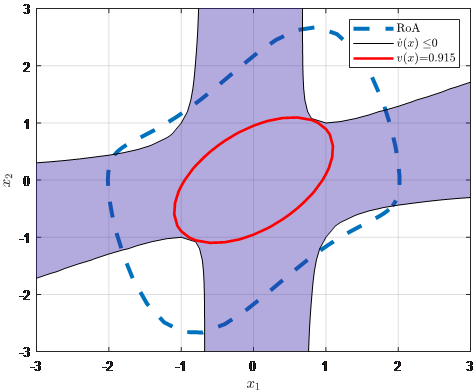}
    \caption{Estimated RoA of the reversed Van der Pol system by energy function constructed in (\ref{eq:12}).}
    \label{fig:RvanderpolROA2}
\end{figure}

The computed matrix \textbf{P} results in the LF (11), whose maximum estimated RoA is presented in Fig. \ref{fig:RvanderpolROA2}. For the LF to be valid, the time derivative should be negative, which is also highlighted. Fig. \ref{fig:RvanderpolROA2} shows that not only $\dot{V}(x)$ should be negative, but also $V(x)\leq c$ is also essential.

\begin{equation}\label{eq:12}
      V(x_1, x_2) =  \begin{bmatrix}
  x_1 & x_2
\end{bmatrix} P \begin{bmatrix}
  x_1 \\ 
  x_2
\end{bmatrix} = x_1^2 - x_1x_2 + x_2^2
\end{equation}

\subsection{Dynamical systems as mappings}
A dynamical system can be thought of as a function that maps the initial to the final conditions after a specific time. Some of the main properties related to the dynamical system stability are given by the following definitions \cite{Khalil}-\cite{ReverseTime}.

\textbf{Definition 1.} \textit{Uniqueness of the solution of a dynamical system -} 
A sufficient condition for the uniqueness of the solution is that the function $f$ should be locally Lipschitz, i.e. it is a continuous function, and its derivative with respect to the state variables is bounded \cite{Khalil}.
\begin{equation}\label{eq:13a}
    \parallel f(t, x) - f(t, y) \parallel \text{ } \leq L \parallel x - y\parallel\\
\end{equation}

It should be noted that this is a weaker condition than the differentiability of the function $f$.

\textbf{Definition 2.} \textit{Boundary preservation in a homeomorphic mapping -}
The continuous function $f:X\rightarrow Y$ is a homeomorphic \cite{Homeomorphisms} if it is bijective and its inverse is also continuous, such that,
\begin{equation}\label{eq:13b}
     f(\partial  \mathbb{A}) = \partial(f(\mathbb{A})), \forall \mathbb{A} \subseteq X\\
\end{equation}
which says that if $f$ is a homeomorphic from \textit{X} to \textbf{Y} and if $\mathbb{A}$ is a subset of X; then the image of the boundary of $\mathbb{A}$ is equal to the boundary of the image of $\mathbb{A}$.

\textbf{Definition 3.} \textit{Reverse-time trajectory -}
If $f$ is Lipschitz, then a unique solution trajectory for each initial condition is guaranteed. Moreover, if the differential equations are solved backwards in time, then the same unique trajectory is traversed \cite{ReverseTime}. This means that, $F(x) = \varphi(T, x)$
where F is an invertible function, in which T is a defined amount of time. Therefore, the response of a dynamical system after a given time for initial conditions chosen from a closed set in $\mathbb{B} \subset \mathbb{R}^n$ will lie in a closed set $\mathbb{D} \subset \mathbb{R}^n$ such that,

\begin{equation}\label{eq:14}
    \partial \mathbb{D} = \partial(F(\mathbb{B})) = F(\partial \mathbb{B})\\
\end{equation}
 
This is quite a useful conclusion which means that simulating a dynamical system numerically for a boundary of initial conditions can give the boundary of the final states, and it is guaranteed that for any initial point inside this boundary, the final response lies in the calculated final boundary.

\subsection{Estimating RoA from reverse-time trajectory}
If the initial point selected is in close proximity to the stable equilibrium point, the uniqueness of the solution guarantees that all points in the reversed trajectory will be attracted to the equilibrium point. Additionally, if the stable equilibrium point is bounded by a limit cycle, it can be identified by reversing the trajectory, but this may require a longer simulation time, as demonstrated in Fig. 3.

\begin{figure}[h]
    \centering
    \includegraphics[width=7.5cm]{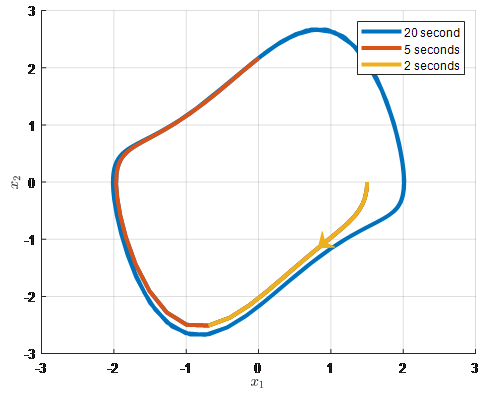}
    \caption{Reverse-time trajectory for the reversed Van der Pol system (5).}
    \label{fig:Traject}
\end{figure}

In power system applications, it is often required that the system reaches a steady state within a specific time frame, known as the settling time \cite{gridCode}. To estimate a time-limited region of attraction (TLRoA) after a disturbance, the system's differential equations are solved backward until the disturbance, assuming a tolerance band (e.g., $\pm$ 5\%) around the equilibrium point. It is crucial that the tolerance band must be a region of attraction. Therefore, a small region of attraction around the equilibrium point is first identified using linearized analysis. Then, the mapping theory explained in Section II-F is used to transform this region into another region through the backward solution of the original ODEs, as shown in Fig.4.

\begin{figure}[h]
    \centering
    \includegraphics[width=7.5cm]{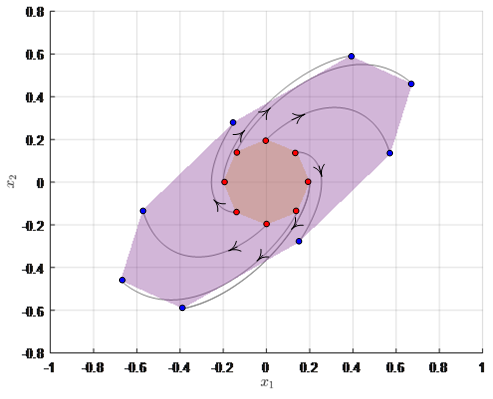}
    \caption{Mapping of the initial states to the final states by the backward solution of the ODEs for the reversed Van der Pol system (5).}
    \label{fig:Traject}
\end{figure}

\section{Transient stability assessment}
\begin{figure*}[h]
    \centering
    \includegraphics[width=15.0cm]{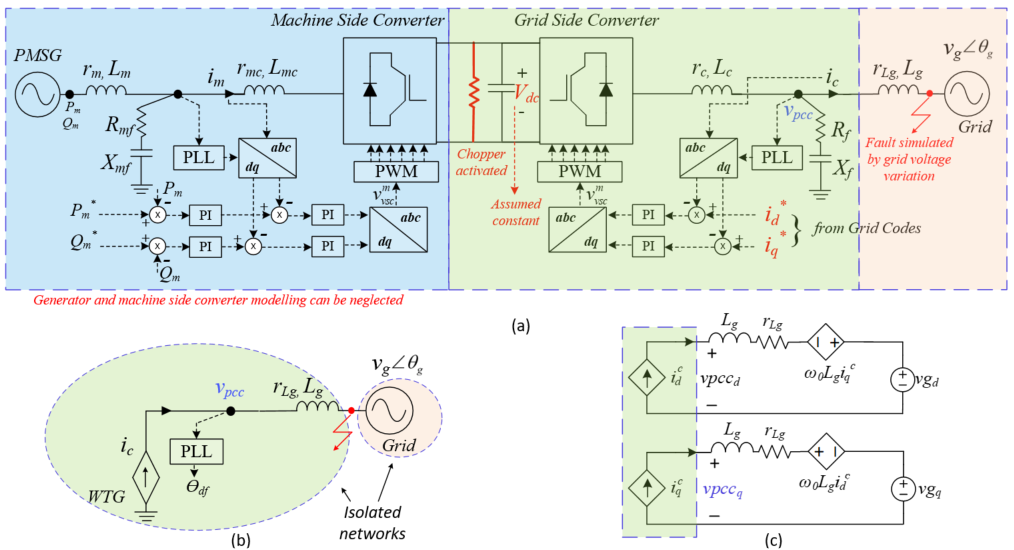}
    \caption{Wind turbine model: (a) Full topology of Type-4 wind turbine system, highlighting the actions/assumptions during faults. (b) Reduced order model (ROM) of the Type-4 wind turbine considering the actions/assumptions. Also showing the synchronisation instability of wind turbine system during grid faults. (c) System representation of ROM in the DQ domain.}
    \label{fig6}
\end{figure*}

Our previous works \cite{18}\cite{33}\cite{34} have demonstrated that a type-4 wind turbine (WT) can be simplified to a grid-side converter with a constant DC voltage during grid faults (Fig. \ref{fig6}a), resulting in a current-controlled source (Fig. \ref{fig6}b) with reference values obtained from the grid codes. For large-signal stability analysis, the fast inner current control dynamics can be neglected, and the shunt capacitor filter's impact on stability can be disregarded if the current is controlled on the grid-side LCL filter. A reduced-order WT model in the DQ domain is presented in Fig. \ref{fig6}c, with SRF PLL for synchronisation.

The equivalent swing equation of the WT converter system derived in \cite{34} can be presented as,

\begin{equation}\label{SEq_1}
M_{eq} \ddot{\delta} = T_{m_{eq}} - T_{e_{eq}} - D_{eq}\dot{\delta}
\end{equation}
where,
\begin{equation}\label{SEq_2}
\begin{aligned}
M_{eq} &= 1- k_p L_g i_d^c\\
T_{m_{eq}} 
&= k_p( \dot{\overline{r_{Lg} i_q^c}} + \ddot{\overline{L_g i_q^c}} + \dot{\overline{L_g i_d^c}} \omega_g)
+ k_i( r_{L_g} i_q^c \\
&\qquad+ \dot{\overline{L_g i_q^c}} + L_g i_d^c \omega_g) \\
T_{e_{eq}} 
&=  (k_i V_g \text{sin} \delta + k_p \dot{V_g} \text{sin} \delta) + M \dot{\omega}_g\\
D_{eq} 
&= k_p ( V_g \text{cos}\delta - \dot{\overline{L_g i_d^c}}) - k_i L_g i_d^c
\end{aligned}
\end{equation}

Equation (\ref{SEq_1}) represents a second-order nonlinear damped differential equation that is used to model the wind turbine (WT) system. This equation takes into account the time-varying nature of the system parameters, which are represented by the derivatives in (\ref{SEq_2}). The WT system is modelled in a DQ frame, rotating at a fixed frequency $\omega_0$. The equation also includes several variables and constants, such as $k_p$ and $k_i$, which are the PLL controller gains, $i_d^c$ and $i_q^c$, which are the converter currents in the converter reference frame, $r_{Lg}$ and $L_g$, which are the grid impedance, and $V_g$ and $\omega_g$, which are the grid voltage and frequency.

Table 1 presents the operating point of the WT converter system considered in this study.

\begin{table}[h]
\caption{SYSTEM AND CONTROL PARAMETERS \cite{36}}
\label{table}
\setlength{\tabcolsep}{3pt}
\begin{tabular}{|p{40pt}|p{105pt}|p{70pt}|}
\hline
Symbol& 
Description& 
Value \\
\hline
$S_b $& 
Rated power& 
12 MVA \\
$V_g$& 
Nominal grid voltage (L-N, pk) & 
690 $\sqrt{2/3}$ V\\
$f_g$& 
Rated frequency & 
50 Hz \\
$r_{Lg}$, $L_{g}$& 
Grid-side impedance & 
SCR=1.2, X/R=18.6 \\
$i_d^c$, $i_q^c$& 
Pre-disturbance active and reactive currents (pu) & 
1.0, 0 \\
$K_{pll}$& 
SRF PLL design: $k_p$ $k_i$ & 
0.025, 1.5 \\
$i_d^{c, ramp}$& 
Post fault active current ramp rate & 
28.4 kA/s and 42.6 kA/s\\
\hline
\multicolumn{3}{p{240pt}}{The PLL gains are chosen to obtain an oscillatory PLL behavior.}
\end{tabular}
\label{tab1}
\end{table}

\subsection{Estimating the TLRoA}
\subsubsection{Lyapunov function candidate from linerised system}
The first step in estimating the WT system boundary is obtaining an initial RoA, which is carried out by constructing a Lyapunov function from the linearised equations of the WT system (15). In \cite{18}, the system (15) was linearised around $\Tilde{\textbf{\textit{x}}}$, which gives
\begin{equation}
\begin{aligned}
\textbf{A} &= \frac{\partial f}{\partial x} |_{x = \Tilde{x}}\\
&=
\left[\begin{matrix}
  0 & 1\\
  \frac{\mp k_i V_g}{1- k_p L_g i_d^c} \sqrt{1-\gamma^2} & \frac{k_i L_g i_d^c \mp k_p V_g \sqrt{1-\gamma^2}}{1- k_p L_g i_d^c} 
\end{matrix}\right] 
\end{aligned}
\end{equation}

\begin{figure}[h]
    \centering
    \includegraphics[width=7.5cm]{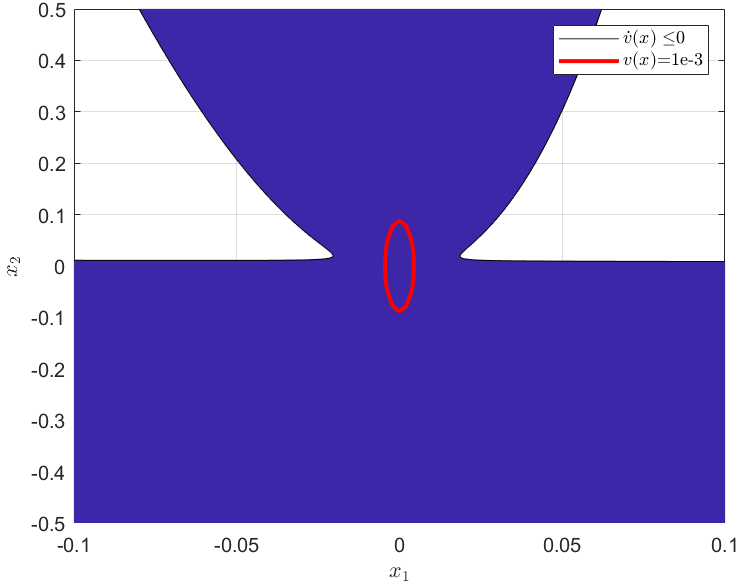}
    \caption{RoA estimate by the linearized model (\ref{SEq_1}) with equilibrium point shifted to the origin.}
    \label{fig:TrajectLRoa}
\end{figure}

By assuming, $\textbf{Q}$ as an identity matrix, $\textbf{P}$ can be computed from (8). Further, similar to (11), the LF constructed from the linear WT system is presented in (\ref{eq:12z}), where Fig. \ref{fig:TrajectLRoa} illustrates the estimated initial RoA, with a selected energy level set of 0.001. It must be noted that for simplicity, in Fig. 6, the equilibrium point was shifted to the origin.

\begin{equation}\label{eq:12z}
      V(\textbf{\textit{x}}) = ax_1^2 +bx_1x_2 + cx_2^2
\end{equation}
where, $a$=49.66, $b$=0.0026, and $c$=0.129.

\subsubsection{Backward solution and trajectory reversing}
\begin{figure}[h]
    \centering
    \includegraphics[width=7.5cm]{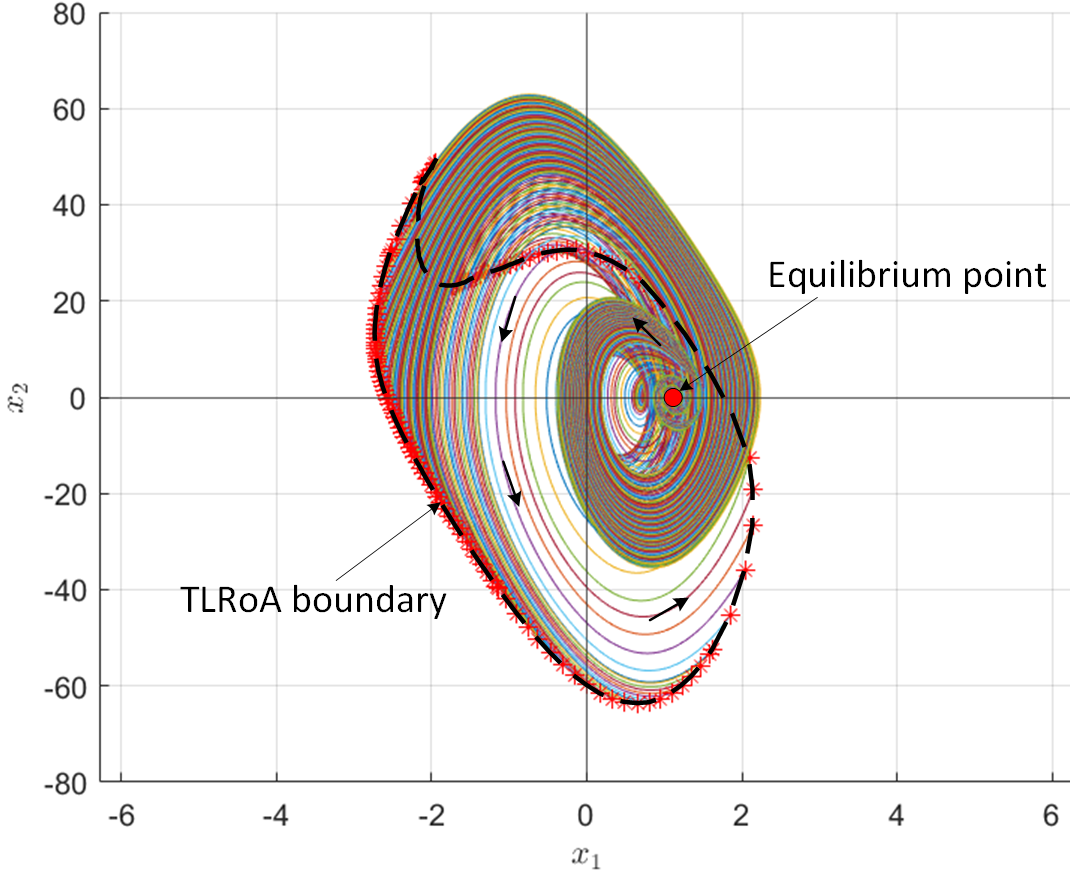}
    \caption{Estimated TLRoA for a post-fault active current ramp rate of 28.4 kA/s.}
    \label{fig:estimated TLRoA}
\end{figure}

In order to estimate the TLRoA, the system (\ref{SEq_1}) can be solved backwards in time using the initial conditions obtained from the initial RoA boundary, which was determined in Fig. \ref{fig:TrajectLRoa}. For this study, the system was solved backwards for 2.25 seconds, which is the time for the oscillations to dampen out. After this period, the system was further solved backwards until the post-fault active current was ramped down to the fault clearing time. The estimated TLRoA with a post-fault active current ramp rate of 28.4 kA/s, as given in (15), is shown in Fig. \ref{fig:estimated TLRoA}. Unlike RoA, for TLRoA all the points at the boundary reaches the equilibrium point at the same time, therefore the enclosed TLRoA is a subset of the actual RoA. 

To generate the smooth TLRoA in Fig. \ref{fig:estimated TLRoA}, the number of samples (initial conditions) are N=186. For a similar system, to generate the actual RoA through forward simulations required N=3213 \cite{33}. This brings out the advantage our proposed method. 
It must be noted that one should ensure enough samples are taken from the boundary of initial RoA to have a smooth boundary for the set of final conditions. This limitation is a known issue in numerical computations, and there are adaptive sampling techniques to reduce the step size, when there are large variations in a function. The methodology to choose the sampling rate is not the focus of this paper, and will be addressed through future publications.

\subsection{Transient stability assessment methodology}
\begin{figure}[h]
    \centering
    \includegraphics[width=7.5cm]{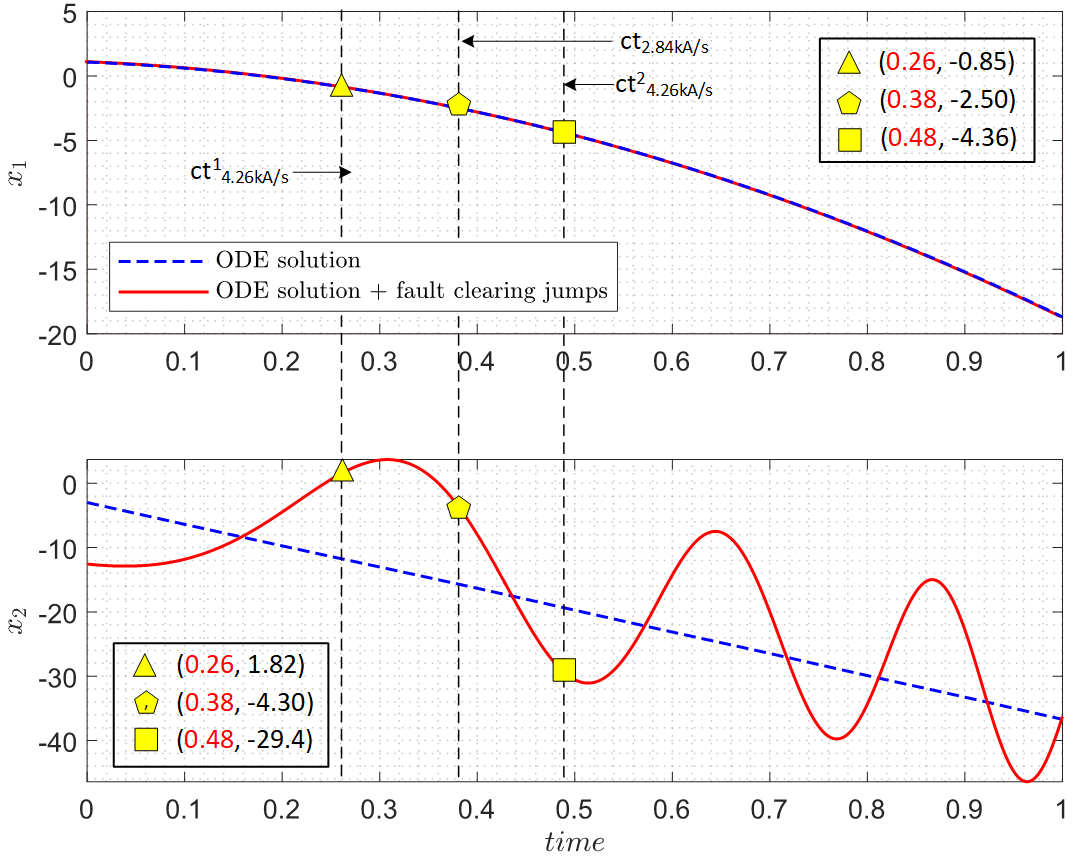}
    \caption{Forward solution of the system during severe fault ($V_g$ = 0 pu, $i_d^c$ = 0.01 pu, and $i_q^c$ = -1 pu).}
    \label{fig:FaultSim}
\end{figure}

The primary question in assessing transient stability is whether a power system can return to equilibrium following a disturbance. To address this, a hybrid approach is proposed, where the post-disturbance system is represented by the estimated TLRoA, and a forward time-domain simulation is carried out to observe the system behaviour during the disturbance. 
Fig. \ref{fig:FaultSim} shows the simulation of a balanced fault (severe grid voltage dip) for an extended period with $V_g=0$ pu, $i_d=0.01$ pu, and $i_q=-1$ pu.
As pointed out in \cite{34}, there will be jumps in the PLL angle and frequency after fault clearance. Therefore, an additional curve (red) is calculated in Fig. \ref{fig:FaultSim} that depicts the PLL angle and frequency with the jumps when the fault is cleared at that time.

\begin{figure}[h]
    \centering
    \includegraphics[width=7.5cm]{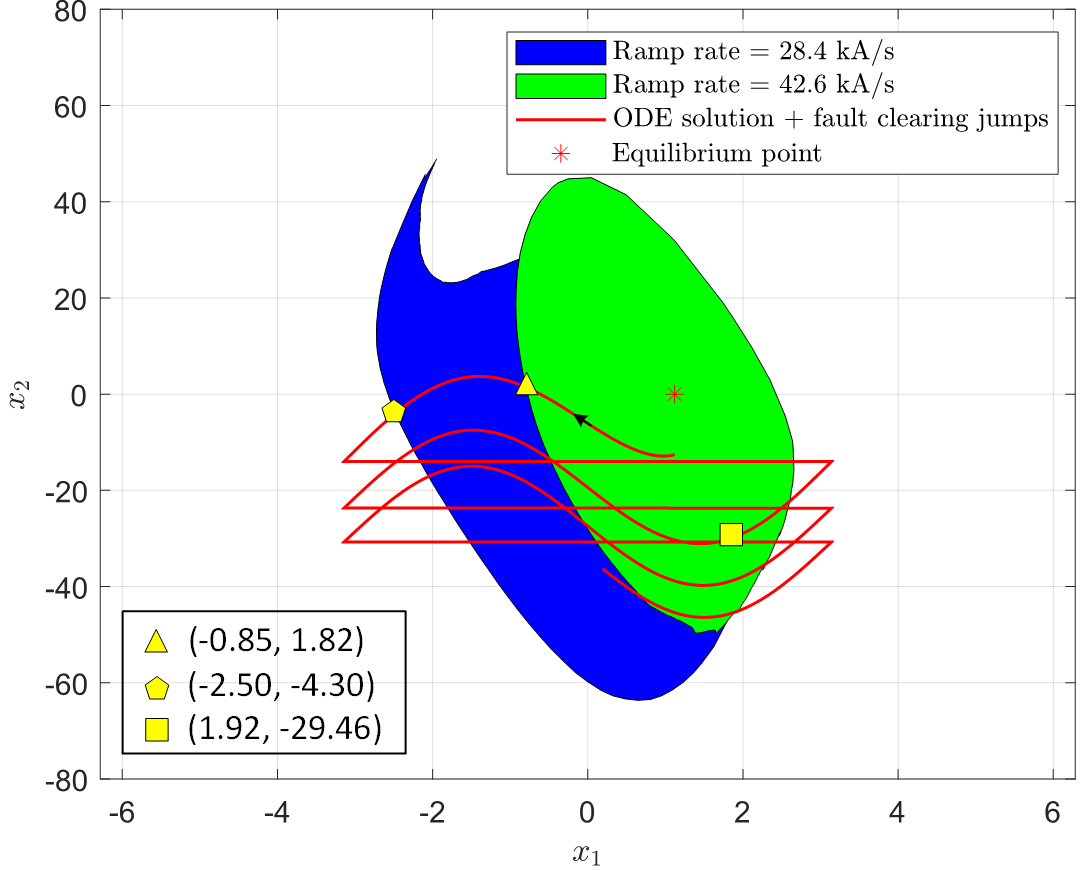}
    \caption{Proposed transient stability assessment method.}
    \label{fig:FaultRoA}
\end{figure}

Based on Section III-A, Fig. \ref{fig:FaultRoA} presents the estimated TLRoA for the system (15) with two different post-fault active current ramp rates 28.4 kA/s and 42.6 kA/s. As expected, the system with a faster ramp rate has a smaller TLRoA \cite{18}. Additionally, the red curve from Fig. \ref{fig:FaultSim} is overlaid on the estimated TLRoA in Fig. \ref{fig:FaultRoA} in delta-omega coordinates, with the PLL angle reset to $\pi$ when it reaches $-\pi$. This is to eliminate the illustration of neighboring ROAs for equilibrium points that repeat every $2\pi$.

For assessing transient stability, it is proposed that clearing the fault at any point along the fault trajectory (red line) inside the TLRoA will guarantee the system's attraction to its post-fault equilibrium point. For instance, if the fault is cleared before reaching the 'yellow triangle' in Fig. \ref{fig:FaultRoA}, then both systems with active current ramp rates of 28.4 kA/s and 42.6 kA/s will be stable. If the fault persists beyond the 'yellow triangle' (but not beyond the 'yellow pentagon'), then only the system with an active current ramp rate of 28.4 kA/s will be stable. Similarly, if the fault persists beyond the 'yellow pentagon' (but not beyond $-\pi$), then both systems will become unstable. Moreover, if the fault persists until the 'yellow square', both systems will be stable again. Thus, it is observed that the fault trajectory exits and re-enters the TLRoA multiple times, suggesting that the WT system can be stable if the fault is cleared at a later time, indicating the WT system has multiple critical clearing times.

The times at which the fault trajectory reaches the critical points, indicated by the 'yellow triangle' and 'yellow pentagon', can be read off from Fig. \ref{fig:FaultSim}, with the x-axis showing the clearing times. The clearing times and the resulting system stability will be later verified using actual EMT WT models.

\section{Time-domain verification}
In this section, the proposed transient stability assessment method is evaluated through time-domain simulations using an EMT WT switching model in PSCAD. The EMT model is designed based on the configuration described in \cite{34}, where the current controller gains are adjusted to achieve a fast response. The system stability against the fault clearing times obtained from Fig. \ref{fig:FaultSim} will be validated.  

\begin{figure}[h]
    \centering
    \includegraphics[width=7.5cm]{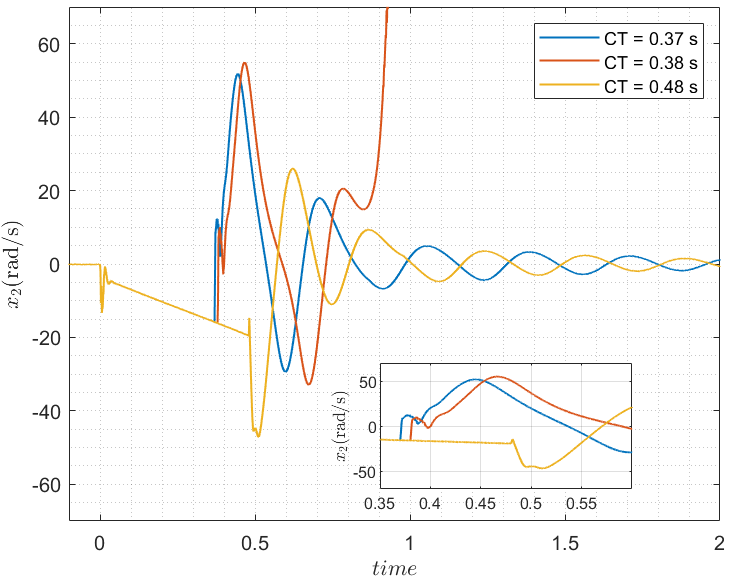}
    \caption{Verification of the identified critical clearing time by PSCAD simulations - recovery ramp rate of 2.86 kA/s.}
    \label{fig:Ramp1}
\end{figure}

\begin{figure}[h]
    \centering
    \includegraphics[width=7.5cm]{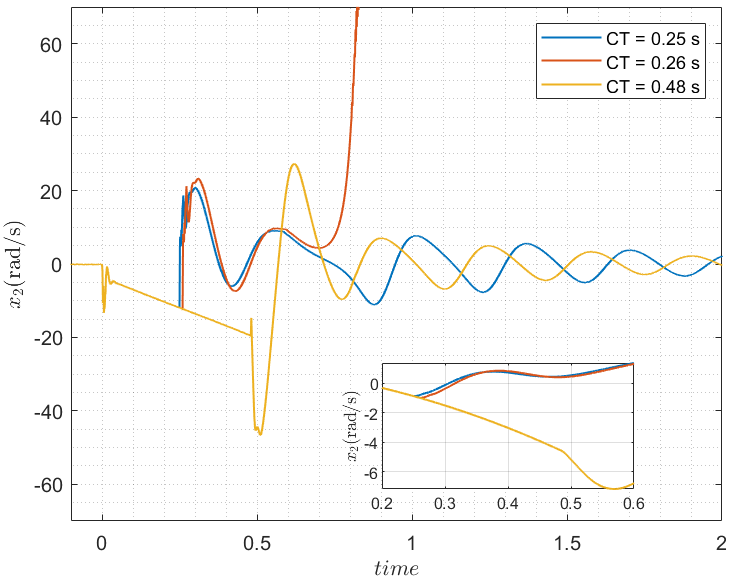}
    \caption{Verification of the identified critical clearing time by PSCAD simulations - recovery ramp rate of 42.6 kA/s.}
    \label{fig:Ramp2}
\end{figure}

Figures \ref{fig:Ramp1} and \ref{fig:Ramp2} show the PSCAD time-domain simulations indicating the clearing time for the WT systems with a post-fault active current ramp rate of 28.4 kA/s and 42.6 kA/s, respectively. It is observed that the clearing times obtained from Fig. \ref{fig:FaultSim} and Fig. \ref{fig:FaultRoA} are consistent with the results obtained from the PSCAD simulations. Additionally, it can be seen that a later clearing time helps the system to regain stability, which is again consistent with the results obtained from our methodology.

Overall, the proposed methodology has a high level of confidence in its application for investigating the transient stability of WTs. The traditional method of estimating the post-fault system RoA through forward-time simulation involves guessing initial conditions, resulting in either a stable or an unstable trajectory. In contrast, our proposed methodology only solves stable trajectories in reverse time, resulting in a quicker and more efficient estimation of the RoA. 

While some efforts have been made to analytically estimate the RoA of a WT with non-autonomous behaviour, such as ramps in active current, these methodologies can be complex and conservative. In contrast, our proposed methodology utilizes a simplified analytical LF to estimate the initial conditions for the reverse time trajectory solutions. As a result, the industry can adopt this methodology without requiring complex mathematical analysis.

Our proposed methodology can enable power system operators and wind farm owners to take advantage of multiple critical clearing times for WT systems. By clearing faults later, uninterrupted supply from the WTs can be achieved, which benefits both parties. This can motivate the development of new power system protection philosophies and smart relays, which can enhance the overall stability and reliability of the power system.

\section{Conclusion}
This work extends our research on nonlinear modelling and transient stability assessment of wind power plants (WPPs) and presents a methodology for transient stability assessment of a WT system with a non-autonomous behaviour. The following are the conclusion of the paper,

\begin{enumerate}
    \item A hybrid approach based on energy function and reverse time-trajectory provides a good estimate of the post-fault system boundary (RoA), where it was observed that the clearing times obtained from the proposed method are consistent with the results obtained from the PSCAD simulations.
    
    \item This work brings out the clear distinction between the system boundaries with different post-fault active current ramp rate controls, i.e. a system with a faster post fault active current ramp rate has a smaller RoA.
    
    \item A new perspective on critical clearing times for wind turbine systems was discussed, which showed that sometimes a later clearing time helps the system to regain stability, motivating the development of new power system protection philosophies and smart relays, which can enhance the overall stability and reliability of the power system.
    
\end{enumerate}


\begin{thebibliography}{00}

\bibitem{1}
GWEC, Global wind report 2021.

\bibitem{2}
J. Tan and Y. Zhang, "Coordinated Control Strategy of a Battery Energy Storage System to Support a Wind Power Plant Providing Multi-Timescale Frequency Ancillary Services," in IEEE Transactions on Sustainable Energy, vol. 8, no. 3, pp. 1140-1153, July 2017, doi: 10.1109/TSTE.2017.2663334.

\bibitem{3}
Y. Huang, X. Yuan, J. Hu and P. Zhou, "Modeling of VSC Connected to Weak Grid for Stability Analysis of DC-Link Voltage Control," in IEEE Journal of Emerging and Selected Topics in Power Electronics, vol. 3, no. 4, pp. 1193-1204, Dec. 2015, doi: 10.1109/JESTPE.2015.2423494.

\bibitem{4}
C. Zhang, X. Cai, A. Rygg and M. Molinas, "Sequence Domain SISO Equivalent Models of a Grid-Tied Voltage Source Converter System for Small-Signal Stability Analysis," in IEEE Transactions on Energy Conversion, vol. 33, no. 2, pp. 741-749, June 2018, doi: 10.1109/TEC.2017.2766217.

\bibitem{7}
J. Hu, Q. Hu, B. Wang, H. Tang and Y. Chi, "Small Signal Instability of PLL-Synchronized Type-4 Wind Turbines Connected to High-Impedance AC Grid During LVRT," in IEEE Transactions on Energy Conversion, vol. 31, no. 4, pp. 1676-1687, Dec. 2016, doi: 10.1109/TEC.2016.2577606.

\bibitem{8}
Wu, F., Zhang, X.-P., Godfrey, K., Ju, P. "Small signal stability analysis and optimal control of a wind turbine with doubly fed induction generator", (2007) IET Generation, Transmission and Distribution, 1 (5), pp. 751-760, doi: 10.1049/ict-gtd:20060395.


\bibitem{9}
H. Liu and J. Sun, "Voltage Stability and Control of Offshore Wind Farms With AC Collection and HVDC Transmission," in IEEE Journal of Emerging and Selected Topics in Power Electronics, vol. 2, no. 4, pp. 1181-1189, Dec. 2014, doi: 10.1109/JESTPE.2014.2361290.

\bibitem{11}
M. Amin and M. Molinas, "Understanding the Origin of Oscillatory Phenomena Observed Between Wind Farms and HVdc Systems," in IEEE Journal of Emerging and Selected Topics in Power Electronics, vol. 5, no. 1, pp. 378-392, March 2017, doi: 10.1109/JESTPE.2016.2620378.

\bibitem{12}
S. Ghosh, G. Yang, M. K. B. Dowlatabadi and L. Kocewiak, "Nonlinear stability analysis of a reduced-order wind turbine VSC-grid model operating in weak grid conditions," 20th International Workshop on Large-Scale Integration of Wind Power into Power Systems as well as on Transmission Networks for Offshore Wind Power Plants (WIW 2021), 2021, pp. 457-463.

\bibitem{13}
M. Bravo, A. Garcés, O. D. Montoya and C. R. Baier, "Nonlinear Analysis for the Three-Phase PLL: A New Look for a Classical Problem," 2018 IEEE 19th Workshop on Control and Modeling for Power Electronics (COMPEL), Padua, Italy, 2018, pp. 1-6, doi: 10.1109/COMPEL.2018.8460081.

\bibitem{14}
S. L. Brunton, B. W. Brunton, J. L. Proctor, and J. N. Kutz, “Koopman invariant subspaces and finite linear representations of nonlinear dynamical systems for control,”PLOS ONE, vol. 11, no. 2, pp. 1–19, Feb. 2016. doi:10.1371/journal.pone.0150171

\bibitem{15}
T. L. Vu and K. Turitsyn, "Lyapunov Functions Family Approach to Transient Stability Assessment," in IEEE Transactions on Power Systems, vol. 31, no. 2, pp. 1269-1277, March 2016, doi: 10.1109/TPWRS.2015.2425885.

\bibitem{16}
Hill, D.J., Hiskens, I.A. and Mareels, I.M.Y. "Stability theory of differential \ algebraic models of power systems," Sadhana 18, 731–747 (1993), doi.org/10.1007/BF03024222

\bibitem{17}
Z. Tian et al., "Hamilton-Based Stability Criterion and Attraction Region Estimation for Grid-Tied Inverters Under Large-Signal Disturbances," in IEEE Journal of Emerging and Selected Topics in Power Electronics, vol. 10, no. 1, pp. 413-423, Feb. 2022, doi: 10.1109/JESTPE.2021.3076189.

\bibitem{18}
Ghosh S., M. K. Bakhshizadeh, Yang G., Kocewiak Ł., Pal B., and Nadarajah M., (2023). "Nonlinear Stability Investigation Of Wind Turbines With Non-autonomous Behavior Based On Transient Damping Characteristics," https://doi.org/10.5281/zenodo.7634543

\bibitem{19}
Baker, John W. “Stability Properties of a Second Order Damped and Forced Nonlinear Differential Equation.” SIAM Journal on Applied Mathematics 27, no. 1 (1974): 159–66. http://www.jstor.org/stable/2100272.

\bibitem{20}
Izumi S., Somekawa H., Xin X., Yamasaki T.,
"Estimation of regions of attraction of power systems by using sum of squares programming", (2018) Electrical Engineering, 100 (4), pp. 2205 - 2216, doi: 10.1007/s00202-018-0690-z



\bibitem{22}
B. Wang, B. Fang, Y. Wang, H. Liu and Y. Liu, "Power System Transient Stability Assessment Based on Big Data and the Core Vector Machine," in IEEE Transactions on Smart Grid, vol. 7, no. 5, pp. 2561-2570, Sept. 2016, doi: 10.1109/TSG.2016.2549063.

\bibitem{24}
Contessa G., "Scientific models and fictional objects",
(2010) Synthese, 172 (2), pp. 215 - 229, doi: 10.1007/s11229-009-9503-2

\bibitem{25}
Nelson R.A., Olsson M.G., "The pendulum—Rich physics from a simple system",v(1986) American Journal of Physics, 54 (2), pp. 112 - 121, doi: 10.1119/1.14703

\bibitem{26}
Furuta K., Iwase M., "Swing-up time analysis of pendulum", (2004) Bulletin of the Polish Academy of Sciences: Technical Sciences, 52 (3), pp. 153 - 163.

\bibitem{27}
Birkhoff G.D., "The restricted problem of three bodies",
(1915) Rendiconti del Circolo Matematico di Palermo, 39 (1), pp. 265 - 334, doi: 10.1007/BF03015982

\bibitem{28}
D.J. Miller,"Realism and time symmetry in quantum mechanics", Physics Letters A, Volume 222, Issues 1–2, 1996, Pages 31-36, ISSN 0375-9601, doi.org/10.1016/0375-9601(96)00620-2.

\bibitem{29}
Aharonov, Yakir, and Jeff Tollaksen. "New insights on time-symmetry in quantum mechanics." arXiv preprint arXiv:0706.1232 (2007).

\bibitem{30}
Vlad,  S.E., "Boolean Functions: Topics in Asynchronicity, 1st ed., Wiley: New York, NY, USA,  2019, doi:10.1002/9781119517528.ch12. 

\bibitem{31}
Lamb, J.; Roberts, J., "Time-reversal symmetry in dynamical systems: A survey", Phys. Nonlinear Phenom 1998, 112, 1–39, doi:10.1016/S0167-2789(97)00199-1.

\bibitem{33}
S. Ghosh,  M. K. Bakhshizadeh, G. Yang, and L. Kocewiak, "Non-linear Stability Boundary Assessment of Offshore Wind Power Plants Under Large Grid Disturbances," 21st International Workshop on Large-Scale Integration of Wind Power into Power Systems as well as on Transmission Networks for Offshore Wind Power Plants (WIW 2022), 2022.

\bibitem{34}
M. K. Bakhshizadeh, S. Ghosh, Ł. Kocewiak, and G. Yang, "Improved Reduced-Order Model for PLL Instability Investigations", (2022), doi.org/10.5281/zenodo.7016303

\bibitem{35}
S. Ghosh, G. Yang, M. K. Bakhshizadeh and L. Kocewiak, "Nonlinear stability analysis of a reduced-order wind turbine VSC-grid model operating in weak grid conditions," 20th International Workshop on Large-Scale Integration of Wind Power into Power Systems as well as on Transmission Networks for Offshore Wind Power Plants (WIW 2021), 2021, pp. 457-463.

\bibitem{Khalil}
Khalil H. K. (1996)., "Nonlinear systems" (2nd ed.), Prentice Hall.

\bibitem{Homeomorphisms}
“The Boundary of a Set under Homeomorphisms on Topological Spaces - Mathonline.” http://mathonline.wikidot.com/the-boundary-of-a-set-under-homeomorphisms-on-topological-sp (accessed Nov. 07, 2022).

\bibitem{ReverseTime}
Roberts, J.; Quispel, G., "Chaos and time-reversal symmetry. Order and chaos in reversible dynamical systems", Phys. Rep.1992,216, 63–177, doi:10.1016/0370-1573(92)90163-T.

\bibitem{gridCode}
“Offshore-Netzanschlussregeln (O-NAR)”, Tech. rep. TenneT TSO GmbH, August 2019.

\bibitem{36}
Ł. Kocewiak, R. Blasco‐Giménez, C. Buchhagen, J. B. Kwon, Y. Sun, A. Schwanka Trevisan, M. Larsson, X. Wang, ``Overview, Status and Outline of Stability Analysis in Converter‐based Power Systems,'' \emph{The 19th International Workshop on Large-Scale Integration of Wind Power into Power Systems.} as well as Transmission Networks for Offshore Wind Farms, Energynautics GmbH, 11-12 November 2020.



































\end{thebibliography}
\end{document}